\documentstyle[11pt]{article}

\newtheorem{lemma}{Lemma}

\title{On Maslov Conjecture about Square Root Type
Singular Solutions of the Shallow Water Equations\thanks{
Supported by RFBR (Russia) grant no.~99--00415 and CONACyT
(Mexico) grant no.~32383-E } }

\author{
S.~Yu.~Dobrokhotov\thanks{
Institute for problems in mechanics, Russian Academy of Sciences,
e-mail: {\tt dobr@ipmnet.ru}}
\and
K.~V.~Pankrashkin\thanks{
Institute for problems in mechanics,
Russian Academy of Sciences and
Institute for mathematics,
Humboldt University at Berlin,
e-mail: {\tt kpankrashkin@mail.ru}}
\and
E.~S.~Semenov\thanks{
Institute of natural sciences and ecology,
Russian research center ``Kurchatov Institute'',
e-mail: {\tt stess@inse.kiae.ru}}
}

\date{}

\begin{document}
\maketitle

\section{Introduction}

About twenty years ago, V.~P.~Maslov~\cite{maslov} put forward
the idea that numerous quasilinear hyperbolic systems have only
finite number of singular solution in general position. These
solutions are shock waves, ``infinitely narrow'' solitons and
point singularities of the type of the square root of a quadratic
form. He has also stated conjecture that such solutions for shallow
water equation can describe the dynamics of mesoscale vortices in
the atmosphere, and trajectories of singularities correspond to
trajectories of these vortices. Interesting integrability
properties of such solutions were found in
works~\cite{dobr1,dobr2,dobr3,dobr4}, where the shallow water
equations on the $\beta$-plane with variable Coriolis
force~\cite{pedl,dolz} were considered :
\begin{equation}
                         \label{e1}
\frac{\partial\eta}{\partial t}+
\langle\nabla,\eta {\bf u}\rangle=0, \quad
\frac{\partial {\bf u}}{\partial t}+
\langle{\bf u},\nabla\rangle{\bf u} -\omega {\bf T}{\bf u}
+\nabla\eta=0.
\end{equation}
Here $x=(x_1,x_2)\in{\bf R}^2$, and the unknowns
are the two-dimensional vector ${\bf u}=({\bf u}_1,{\bf u}_2)$,
and the function $\eta$ that is the geopotential
of the atmosphere or the free surface elevation in
the surface waves theory,
$
{\bf T}=\left(\begin{array}{cc}0 & 1\\ -1 & 0\end{array}\right),
\quad
\nabla=(\frac{\partial}{\partial x_1},\frac{\partial}{\partial x_2}),
$
$\omega=\tilde\omega+\beta x_2$ is the doubled Coriolis
frequency on the $\beta$-plane, and $\tilde\omega$, $\beta$
are some physical constants. The angle brackets stand for
the inner product.

On the some time interval $t\in[0,T]$,
the mentioned singular solutions have the following form:
\begin{equation}
                    \label{e2}
\left(\begin{array}{c}
{\bf u}(x,t)\\
\eta(x,t)
\end{array}\right)
=\left(\begin{array}{c}
u(x,t)\\
\rho(x,t)
\end{array}\right)
+\left(\begin{array}{c}
U(x,t)\\
R(x,t)
\end{array}\right) F(S(x,t)),
\end{equation}
where $F(\tau)=\sqrt{\tau}$.
Here the ``background'' $u=(u_1,u_2)$, $\rho$ and the ``vector amplitude''
$U=(U_1,U_2), R$ are smooth (vector) functions.
The phase function $S$ is a smooth non-negative function
vanishing only on the trajectory $x=X(t)$ of the singularity
and satisfies additionally the following condition:
{\bf Condition (i)}
{\em The matrix ${\bf H}(t)=(\partial^2 S/\partial x_i\partial x_j)
(X(t),t)$
is strictly positive and has distinct eigenvalues.}

Some arguments in favor of the importance of
solutions of the form~(\ref{e2}) are given in~\cite{maslov,
dobr1,dobr2,dobr3,dobr4,zh,maslov2}; these solutions
are elements of some algebra of singular solutions
of quasilinear hyperbolic systems.
The solutions of the form~(\ref{e2}) are selected
from a much wider class of singular solutions
by the general position condition (i).
This class consists of solutions of the form (\ref{e2})
with functions $F$ satisfying the following
conditions:
{\bf Condition (ii)} {\it {\rm a.} $F(\tau)$ is a continuous functions
as $\tau\ge0$, $F(0)=0$; {\rm b.} $F(\tau)$ is smooth for $\tau>0$,
$\lim_{\tau\to+0}F'(\tau)=\infty$.}

Obviously, the function $F$ in~(\ref{e2}) is not
determined uniquely by the condition (ii):
we can multiply it on any non-vanishing smooth
function and add any smooth function vanishing on the trajectory
$X(t)$. Moreover, the function $S$ can be also multiplied by
non-vanishing smooth function.

As was noted, it was announced in~\cite{maslov},
that for numerous hyperbolic systems the condition~(i)
"kills" (modulo mentioned possible changes in (2)) all the solutions~(\ref{e2}) excepting
$F=\sqrt{\tau}$  in the sense that if $F\not=\sqrt{\tau}$,
then all the $x$-derivatives of $U$ and $R$
are zero on the trajectory $X(t)$.
The proof of this conjecture was given first
by V.~N.~Zhikharev~\cite{zh}, but his arguments
were not complete even they contained many elegant constructions.
It is impossible to rectify his proof, because
it needs really considering of the problem from the
very beginning.

Our aim is to give a
complete proof of the Maslov conjecture for equations~(\ref{e1}).
There is no enough place in this paper for the whole proof,
so here we only give the general scheme,
accent the  principal moments and
the ideas,  formulate basic main and auxiliary statements,
and  illustrate  proofs for some of them.
The complete proof will appear in ~\cite{full}.
Our approach is based on algebraic construction rather than
to them related to differential equations, and it seems,
that the similar considerations are valid for much wider
class of hyperbolic quasilinear system.

\smallskip

\noindent{\bf Theorem}
{\it Suppose that the functions specifying the solution~{\rm(\ref{e2})}
satisfy the conditions {\rm(i)} and {\rm(ii)}, and some $x$-derivatives
of $U$ and $R$ are non-zero. Then without loss of generality
we can assume $F=\sqrt{\tau}$.}

\section{Auxiliary constructions}

First of all, we pass to the moving coordinate system $(x',t)$
with $x'=x-X(t)$, then system~(\ref{e1}) is written as
(we omit the prime on the new $x$ variable and
put $V=\dot x$, $v={\bf u}-V$)
$$
\label {e3}
\eta_t+\langle\nabla,\eta v\rangle=0,
\quad
v_t+\langle v,\nabla\rangle v+\nabla\eta+\dot V-\omega {\bf T}(v+V)=0.
$$
Substitute now function~(\ref{e2}) into this system,
then we obtain the system
\begin{equation}
                         \label{e4}
F'(\tau)A+F(\tau)B+F(\tau)F'(\tau)C+F^2(\tau)D+E=0.
\end{equation}
Here
$$
\begin{array}{c}
A =\left(\begin{array}{c}
\Lambda R+\rho \langle U,P\rangle\\
R P +\Lambda U
\end{array}\right),\,
B =\left(\begin{array}{c}
R_t+\langle\nabla,\rho U+Ru\rangle\\
\Gamma(U)+\langle U,\nabla\rangle u+\nabla R-w{\bf T}U
\end{array}\right),\,
C =\left(\begin{array}{c}
2R\langle U,P\rangle,\\
\langle U,P\rangle U
\end{array}\right), \\
D =\left(\begin{array}{c}
\langle\nabla,RU\rangle\\
\dot U
\end{array}\right), ~
E =\left(\begin{array}{c}
\rho_t+\langle\nabla,\rho u\rangle\\
\Gamma(u)+\nabla\rho+\dot V
-\omega{\bf T}(u+V).
\end{array}\right),
\end{array}
$$
where $\tau=S(x,t)$, $P=\nabla S$, $\Lambda=S_t+\langle u,P\rangle$, $\dot U=\langle U,\nabla\rangle U$, $\Gamma=\partial/\partial t+\langle u,\nabla\rangle$.
Since the function $S$ can be represented in the form (see
condition (i)) $S=\frac 1 2 \langle x, {\bf H}(t) x\rangle+ O(|x|^3)$,
by using the Morse lemma~\cite{morse} there exists a smooth change of variables
$x=x(y,t)$ such that $S=y^2$. The vectors $A$, $B$,
$C$, $D$, and $E$ depend smoothly on $y$.
We can also introduce polar coordinates $\tau$, $\varphi$,
$
y_1=\sqrt{\tau}\cos\varphi, \quad
y_2=\sqrt{\tau}\sin\varphi,
$
and substitute them into system~(\ref{e4}), then
for each $\varphi$ we obtain a system of first order
differential equations with coefficients depending
smoothly on $\sqrt{\tau}$, $\varphi$, e.c. $A=A(x(y(\sqrt{\tau},\varphi)))$.
To simplify the notation we do not write all these arguments and keep only dependence of coefficients
on $x$.

We will intensively exploit the fact that {\em $F$ does not
depend on $\varphi$}.

Our proof is divided now into three main parts:
\begin{itemize}
\item to obtain from (\ref{e4}) more simple equations
admitting exact formulas for their solutions,
\item to select from these solutions only
those satisfying (ii);
\item to show inconsistency of system (\ref{e4})
corresponding to all the obtained solutions except $F=\sqrt{\tau}$.
\end{itemize}

\section{Model equations}

Since we have a system of three equations
for a single function $F$, one can try to eliminate
terms with $F'$ and $FF'$ and to obtain a quadratic equation.
Unfortunately, because of degeneracy,
it is not always possible, and we have to find another
way of simplification for equations $(\ref{e4})$.

We set $P_\perp={\bf T} P$, $U_\perp={\bf T} U.$
Let  $G(x,t)$ be a smooth scalar or vector function. Then to $G$ we
can assign the Taylor expansion at $x=0$:
$$
G\sim\sum^\infty_{n=0}G^{(n)}(x,t),
$$
where $G^{(n)}(x,t)$
is a homogeneous polynomial of $x$ of degree $n$.
By $o(|x|^\infty)$ we denote smooth functions vanishing to the infinite
order at  $x=0$.

\begin{lemma}[On model equations]\label{111}
Let the condition {\rm(i)} is fulfilled and system~{\rm(\ref{e4})}
has a solution $F$ satisfying condition {\rm(ii)}, then\\[\smallskipamount]
{\bf A.}
there exist smooth functions $\alpha$, $\beta$,
$\gamma$, $\delta$ of $(x,t)$ such that
$F$ satisfies the Riccati equation
\begin{equation}              \label{riccati}
\alpha(x,t)F'(\tau)+\beta(x,t)F^2(\tau)+
\gamma(x,t)F(\tau)+\delta(x,t)=0,
\end{equation}
where  $x$-derivatives
of the function $\alpha$ do not vanish at $x=0$.\\[\medskipamount]
{\bf B.} $F$ satisfies one of the following
three equations:\\[\smallskipamount]
-~the quadratic equation
\begin{equation}             \label{qe}
a(x,t)F^2(\tau)+b(x,t)F(\tau)+c(x,t)=0,
\end{equation}
if $\langle U,P_\perp\rangle\not =o(|x|^\infty)$,
$\langle \dot U,U_\perp\rangle\not=o(|x|^\infty)$, and
$\langle U,P\rangle=o(|x|^\infty)$, or $\langle U,P_\perp\rangle=o(|x|^\infty)$,
\\[\smallskipamount]
-~the linear differential equation
\begin{equation}             \label{lde}
a(x,t)F'(\tau)+b(x,t)F(\tau)+c(x,t)=0,
\end{equation}
if $\langle U,P_\perp\rangle\not=o(|x|^\infty)$,
$\langle \dot U,U_\perp\rangle\not=o(|x|^\infty)$,\\[\smallskipamount]
-~the cubic equation
\begin{equation}              \label{ce}
a(x,t)F^3(\tau)+b(x,t)F^2(\tau)+c(x,t)F(\tau)+d(x,t)=0,
\end{equation}
if $\langle U,P_\perp\rangle\not=o(|x|^\infty)$,
$\langle \dot U, U_\perp\rangle\not=o(|x|^\infty)$,
$\langle U,P\rangle\not=o(|x|^\infty)$.\\[\smallskipamount]
Here also  $a$, $b$, $c$, $d$
are smooth functions and some $x$-derivatives
of $a$ do not vanish at the point $x=0$.\\[\smallskipamount]
{\bf C.} if $\psi=o(|x|^\infty)$, then
$F^n \partial^mF/\partial \tau^m \psi=o(|x|^\infty)$
for any integer $n$ and integer $m\ge0$.
\end{lemma}

Prior to proving lemma \ref{111}, we prove a useful auxiliary assertion.
\begin{lemma}\label{imp}
Let
a smooth vector function $z$ satisfy $\langle z,P\rangle=0$,
then there exists a smooth function $\alpha$ such that
$z=\alpha P_\perp$. It follows from here, that if
$z=z^{(k)}$ is
$k$-st-order homogeneous polynomial in $x$, $z^{(k)}\ne 0$,
then the same assertion is valid with
$z$ replaced by $z^{(k)}$, $P$ replaced by $P^{(1)}$, and
$\alpha$ replaced by $(k-1)$-st-order homogeneous polynomial $\alpha^{(k-1)}.$
\end{lemma}

Obviously, it suffices to prove the assertion pertaining to
$z$. We pass from the coordinates  $x$  to the Morse coordinates $y$.
Then we have
$P={}^t\big(\partial x/\partial y\big)^{-1}\, {}^t(y_1, y_2)$. Consequently,
$\langle ({\partial x}/{\partial y})^{-1}w,y\rangle =0$.
We set  $w=(\partial x/\partial y)^{-1}z$ and represent the components
$w_1$ and $w_2$ of the vector  $w$ in the form
$w_1=w_1^0(y_1,t)+y_2\tilde w_1(y,t)$ and $w_2=w_2^0(y_2,t)+y_1\tilde
w_2(y,t)$, where  $w_i^0$ and $\tilde w_i$ are smooth functions. Then
the orthogonality condition for  $z$ and  $P$  acquires the form
$y_1 w_1^0(y_1,t)+y_2 w_2^0(y_2,t)+y_1 y_2(\tilde w_1+\tilde w_2)=0$.
In this equation, we in turn set $y_1=0$ and  $y_2=0$, thus obtaining
$w_1^0=w_2^0=0$. Next, we divide it by  $y_1y_2$ and obtain
 $\tilde w_1=-\tilde w_2\equiv \tilde g(y,t)$ and
$(\partial x/\partial y)^{-1}z=gTY$. Now note that $T^tQ=\mathop{\rm det}
QQ^{-1}T$ for any non-degenerate $2\times2$ matrix $Q$. Therefore,
$$
z=g\Big(\frac {\partial x}{\partial y}\Big)Ty
=g\mathop{\rm det}\Big(\frac {\partial x}{\partial y}\Big)T^t
\Big(\frac{\partial x}{\partial y}\Big)^{-1}y
=g\Big(\frac {\partial x}{\partial y}\Big)P_\perp.
$$
The proof of the lemma \ref{imp} is complete.

To prove the lemma \ref{111}, we use mostly pure algebraic
procedures. Two cases are studied separately:
$$
(a) \quad  \langle U,P_\perp\rangle\not=o(|x|^\infty),\quad
(b) \quad \langle U,P_\perp\rangle=o(|x|^\infty).
$$
Study first the case $(a)$.
Let us transforms the system (\ref{e4}). The coefficients
of the transformed systems will be indicated by primes;
we do not write system themselves, but only their coefficients,
which are again referred to (\ref{e4}).

Let us multiply the first equation of (\ref{e4})
by $U$ in the sense of the inner product,
and the second and the third equation
by $2R$; then we subtract the first equation from the second and
the third ones; this leads to the equations
with coefficients
$$
A'=\left(\begin{array}{c}
\Lambda R +\rho \langle U,R\rangle\\
2R^2P+(\Lambda R-\rho\langle U,P\rangle)U
\end{array}\right), ~
C'=\left(\begin{array}{c}
2R\langle U,P\rangle\\
0
\end{array}\right), ~
D'=\left(\begin{array}{c}
\langle\nabla,RU\rangle\\
2R\dot U-\langle \nabla,RU\rangle U
\end{array}\right).
$$
In the primed system, we multiply the second and the third equations
by $U_\perp$ and $P_\perp$, this results in system (\ref{e4})
with coefficients
$$
\begin{array}{c}
A''=\left(\begin{array}{c}
\Lambda R+\rho \langle U,P\rangle\\
2R^2\langle U_\perp,P\rangle\\
(\rho\langle U,P\rangle-\Lambda R)
\langle U,P_\perp\rangle
\end{array}\right), \quad
C''=\left(\begin{array}{c}
2R\langle U,P\rangle\\
0\\
0
\end{array}\right),\\
D''=\left(\begin{array}{c}
\langle\nabla, RU\rangle\\
2R\langle\dot U,U_\perp\rangle\\
2R\langle \dot U,P_\perp\rangle-
\langle\nabla ,RU\rangle \langle U,P_\perp\rangle
\end{array}\right).
\end{array}
$$
The second equation in the obtained system is the requested
Riccati equation (\ref{riccati}). The conclusion C.
of the lemma follows directly from the Riccati equation.

Consider now the following sub-cases:
$$
(a.1) \quad \langle \dot U, U_\perp\rangle=o(|x|^\infty),\quad
(a.2) \quad \langle \dot U, U_\perp\rangle\not=o(|x|^\infty).
$$
In case $(a.1)$ the term $\langle \dot U, U_\perp\rangle F^2$
in the obtained Riccati equation can be included into the coefficient
$E''_2$, and we arrive at the equation (\ref{lde}).

Now proceed with the case $(a.2)$. Suppose first
$(a.2.1) \quad \langle U,P\rangle\not= o(|x|^\infty)$.
Let us express $F'$ through $F^2$ and $F$ from
the Riccati expression and
substitute this expression into the first equation.
This results in the cubic equation (\ref{ce}) with non-degenerate
coefficient at $F^3$.
If
$(a.2.2) \quad \langle U,P\rangle= o(|x|^\infty)$, we
multiply the first equation of the double primed system by
$\langle U_\perp, P\rangle$ and add it to
the third equation of the same system; as a result the coefficients
of $F'$ and $FF'$ in the obtained equation are
$o(|x|^\infty)$, and these terms can be included
into the corresponding $E$-coefficient.
The coefficient of $F^2$ is $D'''_3=2(R\langle\dot U,P_\perp\rangle-
\langle\nabla,RU\rangle\langle U,P_\perp\rangle)$.
Under lemma \ref{imp}, $U$ can be represented in the form
$U=\alpha P_\perp+\beta P$, where $\alpha$, $\beta$
are smooth function, $\beta=o(|x|^\infty)$,
and some derivatives of $\alpha$ do not vanish
at $x=0$.
Some calculations reduce expression for $D'''_3$
to the form $2\alpha^2\langle P,{\bf Q} P\rangle+\zeta$, where the matrix ${\bf Q}$
is given by
$$
{\bf Q}=\left(\begin{array}{cc}\displaystyle
-R \frac{\partial^2 S}{\partial x_1\partial x_2} -\langle P_\perp,\nabla R\rangle&
\displaystyle R\frac{\partial^2 S}{\partial x_1^2}\\
\displaystyle R\frac{\partial^2 S}{\partial x_2^2} &
\displaystyle -R \frac{\partial^2 S}{\partial x_2\partial x_1} -\langle P_\perp,\nabla R\rangle
\end{array}\right),
$$
and $\zeta=o(|x|^\infty)$.
Suppose now that $D'''_3=o(|x|^\infty)$, then by virtue of Lemma~\ref{imp}
we obtain ${\bf Q}P=\gamma P+\delta P_\perp$,
where $\gamma$, $\delta$ are smooth functions, and $\gamma=o(|x|^\infty)$.
This means, in particular, that up to $o(|x|^\infty)$
we have $S''_{x_1 x_1}=-S''_{x_2 x_2}$. But in this case  the matrix
${\bf H}$ (see condition (i)) is not positive definite.
This contradiction finishes the proof for case $(a)$.

The consideration of the case $(b)$
is based on similar ideas, but needs more sophisticated calculation.

\section{Possible solutions to the model equations}

Our previous consideration shows that
coefficients of equations (\ref{e4}) are
smooth functions of $\sqrt{\tau}$ only.
Nevertheless, using the assumption about the existence of solutions
satisfying (ii), we show that we can really
suppose them smooth on $\tau$.

The following assertion plays an important role
in all our considerations:
\begin{lemma}
                       \label{lem1}
Let $u$, $v$ be smooth functions of $y$
and some derivatives of $v$ at $y=0$ do not
vanish. If a function $\Phi$ satisfies
$\Phi(\tau)=u(y_1,y_2)/v(y_1,y_2)$, then
$\Phi(\tau)=\tau^n\Psi(\tau)$, where
$\Psi$ is a smooth function, $n$ is an integer number.
\end{lemma}

For the proof, we set
$y_1=\sqrt{\tau}\cos\varphi$,$y_2=\sqrt{\tau}\sin\varphi$,
$\alpha(\tau,\varphi)=$
$u(\sqrt{\tau}\cos\varphi,\sqrt{\tau}\sin\varphi),\quad$
$\beta(\tau,\varphi)=$$v(\sqrt{\tau}\cos\varphi,\sqrt{\tau}\sin\varphi)$,
then $\alpha$ and $\beta$ are smooth functions of $\sqrt{\tau}$,
and $f(\tau)=\frac{\alpha(\tau,\varphi)}{\beta(\tau,\varphi)}$.
Let us extract the leading terms of $\alpha$ and $\beta$:
\begin{equation}
         \label{m-f}
\alpha=\alpha_m(\varphi)\tau^{m/2}+a_m(\tau,\varphi)\tau^{m/2},\quad
\beta=\beta_k(\varphi)\tau^{k/2}+b_k(\tau,\varphi)\tau^{k/2},\quad
a_m(0,\varphi),\,b_k(0,\varphi)=0.
\end{equation}
If $p=m-k$ is odd, then at least in some sector the leading term of $f$
is $\frac{\alpha_m(\varphi)}{\beta_k(\varphi)}\tau^{p/2}$.
Note that $\alpha_m$ and $\beta_k$ can be represented as homogeneous
polynomials of order $m$ and $k$, respectively, in $\cos\varphi$ and
$\sin\varphi$, and one can readily see that the expression (\ref{m-f})
in this case (odd $p$) depends on $\varphi$. Therefore, the number $p$ must be
even. Then we can write
\begin{equation}
     \label{Psi}
\Psi=\frac{W}{Z},
~\mbox{where}~
\Psi(\tau)=\frac{f(\tau)}{\tau^n},\quad
W(\tau,\varphi)=\alpha(\tau,\varphi)\tau^{-m/2},\quad
Z(\tau,\varphi)=\beta(\tau,\varphi)\tau^{-k/2}.
\end{equation}
It is obvious that
$W(0,\varphi),Z(0,\varphi)\ne 0$, and therefore, $\Psi$ is a smooth
function of $\sqrt{\tau}$ and satisfies $\Psi(0)\ne 0$. For brevity, we
set $\xi=\sqrt{\tau}$. Now for each $N$ we can write
$\Psi=\sum_{l=0}^N\Psi_l\xi^l+o(\xi^N)$,
$W=\sum_{l=0}^N W_l(\varphi)\xi^l+o(\xi^{N})$,
$Z=\sum_{l=0}^N Z_l(\varphi)\xi^l+o(\xi^N)$.
Let us substitute these sums into (\ref{Psi}) and match the coefficients of
$\xi^l$; then for $l\le N$ we obtain
$W_l=\sum_{j=0}^l \Psi_j B_{l-j}.$
By taking $N$ sufficiently large, we can obtain the last expression
for every $l$.
These equations for odd $l$ read as follows:
\begin{eqnarray*}
W_1  =\Psi_0 Z_1 + \Psi_1 Z_0,\quad
W_3  =\Psi_0 Z_3 + \Psi_1 Z_2 + \Psi_2 Z_1 + \Psi_3 Z_0,\\
W_5  =\Psi_0 Z_5 + \Psi_1 Z_4 + \Psi_2 Z_3 + \Psi_3 Z_2 +
\Psi_4 Z_1 + \Psi_5 Z_0, \ldots
\end{eqnarray*}
From the first of these equations, we obtain $\Psi_1=0$ (indeed,
$W_1-\Psi_0Z_1$ is a polynomial of odd order in $\cos\phi$ and $\sin\phi$
and cannot be equal to the even-order polynomial $\Psi_1 Z_0$). By
substituting this into the second equation, we obtain $\Psi_3=0$, and so
on. Finally, we see that $\Psi_l=0$ for all odd $l$. Recall that
$\Psi_l=d^l\Psi/d\xi^l(0)$, and there obviously exists a smooth
function $\Phi$ such that $\Psi(\xi)=\Phi(\tau)$. Now we only need to
recall that $\Psi=\tau^{-n}f$.

\medskip

In what follows we use differentiation by $\varphi$:
$\frac{\partial z}{\partial\varphi}=
-y_2\frac{\partial z}{\partial y_1}
+y_1\frac{\partial z}{\partial y_2}$.
The important fact is that if $z$ is also smooth,
then  $z_\varphi$ is also smooth.

\begin{lemma}
                       \label{lem2}
Let $u$, $v$ be smooth functions of $y$ and
some derivatives of $v$ at the point $y=0$
do not vanish. If all the derivatives
of the function $u'_\varphi v-u v'_\varphi$
vanish at $y=0$, then at least in some sector
$\varphi_1\le \varphi\le\varphi_2$ we have
$u=\tau^n\alpha v$, where $n$ is an integer number,
$\alpha$ is a smooth function of $\tau$
\end{lemma}

Let us consider quadratic equation (\ref{qe})
and linear differential equation (\ref{lde}).
Divide both them by $a$ and differentiate by $\varphi$,
then in both cases we obtain a linear equation
$uF+v=0$, where $u=b'_\varphi a-b a'_\varphi$,
$v=c'_\varphi-c a'_\varphi$. If $u$ has non-vanishing
derivatives at $y=0$, then Lemma~\ref{lem1}
and condition (iia) imply smoothness $F$, and, therefore,
$F$ does not satisfy the condition (iib).
Therefore, all the derivatives of both $u$ and $v$
vanish at $y=0$. By applying Lemma~\ref{lem2}
we obtain
$b=\tau^{-n}\beta a$, $c=\tau^{-n} \gamma a$,
where $n$ is a non-negative integer number and $\beta$, $\gamma$
are smooth functions of $\tau$. Dividing now
equations (\ref{qe}), (\ref{lde}) by $a$, we obtain
the equations of the same type:
\begin{eqnarray}
\tau^n F^2+\beta F +\gamma=0, \label{qe-s} \\
\tau^n F'+\beta F+\gamma=0 \label{lde-s},
\end{eqnarray}
but with coefficients depending on $\tau$ smoothly.

Now consider the cubic equation (\ref{ce}).
Divide it also by $a$ and differentiate by $\varphi$,
then we obtain a quadratic equation
$uF^2+vF+w=0$, where $u=b'_\varphi a-b a'_\varphi$,
$v=c'_\varphi-c a'_\varphi$, $w=d'_\varphi a-d a'_\varphi$.
If $u$ has non-vanishing derivatives at $y=0$,
then we have a quadratic equation, which is already studied.
If all the derivatives of $u$ at $y=0$ vanish,
we can include the term $uF^2$ into $w$,
and obtain a linear equation $vF+W$, $W=w+vF^2$.
Again by using Lemma~\ref{lem1} and condition (ii)
we show that all the derivatives of both $v$ and
$W$ (and, consequently, $w$) vanish at $y=0$.
We have again
$b=\tau^{-n}\beta a$,
$c=\tau^{-n} \gamma a$, $d=\tau^{-n} \delta a$,
where $n$ is a non-negative integer number
and $\beta$, $\gamma$, $\delta$ are smooth
functions of $\tau$. Dividing (\ref{ce})
by $a$ and multiplying by $\tau^n$ we reduce
it to a cubic equation with smooth on $\tau$
coefficients:
\begin{equation}  \label{ce-s}
\tau^n F^3+\beta F^2+\gamma F+\delta=0, \quad \alpha=\tau^n.
\end{equation}

\medskip

Now we can use exact formulas for solutions of
obtained equations.

\medskip

For {\bf quadratic equation}~(\ref{qe-s}) we have
$F=(-\beta\pm\sqrt{\beta^2-4\gamma\tau^n})/(2\tau^n)$. If
$\beta^2-4\gamma\tau^n\not=o(|x|^\infty)$, then condition (ii)
immediately gives us $F=\sqrt{\tau}$. If
$\psi=\beta^2-4\gamma\tau^n=o(|x|^\infty)$, then condition (iia)
gives us at least the representation $F=f+\sqrt{\psi}$ with a
smooth function $f$. Let us substitute this representation into
Riccati equation~(\ref{riccati}), than we obtain for
$\sqrt{\psi}$ a Riccati equation with smooth coefficients. This
means, that $\sqrt{\psi}$ is a smooth function, and $F$ does not
satisfy condition (iib).

\medskip

Now let us consider {\bf linear equation} (\ref{lde-s}).
Consider different cases.

Suppose  that $\beta=o(|y|^\infty)$, then the function
$\delta=\beta F+\gamma$ is smooth, $F'=-\delta/\tau^n$, and
condition (iia) implies smoothness of $F$.
Let  $\gamma=o(|y|^\infty)$, then $\delta=\gamma/F$ is smooth,
and we have $\tau^nF'/F+e=0$, where $e=\beta+\delta$ is a smooth function.
Consequently, $\frac{F'(\tau)}{F(\tau)}=-\frac{e(y_1,y_2)}{\tau^n}$,
or by Lemma~\ref{lem1}, $\frac{F'(\tau)}{F(\tau)}=\tau^m\Phi(\tau)$,
where $\Phi$ is a smooth function. Then
$F(\tau)=\exp \int \tau^m\Phi(\tau)d\tau$. Without loss of generality
we can suppose $\Phi(0)\ne 0$. If $m\ge0$, then $F(0)\ne=0$.
If $m<-1$, then either $F(0)=\infty$ or $F'(0)=0$.
Suppose $m=-1$, then $F(\tau)=\tau^\kappa\Psi(y)$, where $\kappa=\Phi(0)$,
$\Psi$ is a smooth function non-vanishing at $y=0$.
The condition (ii) is satisfied if and only if $0<\kappa<1$.

Now suppose that both $\beta$ and $\gamma$ have non-vanishing derivatives.
Let us rewrite our equation in form $F'=AF+B$, where
$A=\tau^m\beta_0$, $B=\tau^k\gamma_0$, where $m$, $k$
are integers, $\beta_0$, $\gamma_0$ are smooth functions
non-vanishing at $y=0$, then $F$ can be written in a form
$$
F(\tau)=\left(\int B(\tau)
\exp\Big(-\int A(\tau)d\tau\Big)d\tau\right)
\exp\Big(\int A(\tau)d\tau\Big).
$$
Suppose {\bf(a)} $m\ge0$, then $\phi_0=\exp\int\alpha(\tau)\,d\tau$
is smooth and $\phi_0(0)\ne 0$. Also
$$
\int B(\tau)\exp\Big(-\int A(\tau)\,d\tau\Big)d\tau=
\int \frac{B(\tau)}{\phi_0(\tau)}\,d\tau= \tau^p\psi_1(\tau)+q\log \tau+
\psi_2(\tau)
$$
for some smooth functions $\psi_1(\tau)$ and $\psi_2(\tau)$, $p\le0$,
$q\in{\bf R}$, and
$F(\tau)=q\phi_0(\tau)\log \tau+\tau^p\phi_1(\tau) +\phi_2(\tau)$,
where $\phi_1$ and $\phi_2$ are smooth functions. One can readily see
that condition~(iia) implies $q=0$; therefore, $F$ cannot satisfy~(iib).

Now suppose {\bf(b)} $m=-1$, then
$\int A(\tau)\,d\tau=\kappa\log \tau+\psi(\tau)$,
for some $\kappa\ne 0$ and some smooth function $\psi$,
and, consequently, $\phi_0=\exp\int a(\tau)\,d\tau=\tau^\kappa \phi_1(\tau)$
for some smooth function $\phi_1(\tau)$ with $\phi_1(0)\ne 0$.
Suppose first that $\kappa$ is integer, then
$$
\int\frac{B(\tau)}{\phi_0(\tau)}\,d\tau= \tau^p\psi_1(\tau)+Z\log
\tau +\psi_2(\tau),
$$
where  $p\le0$, $\psi_1$ and $\psi_2$ are smooth functions, $Z\in{\bf R}$,
and $F(\tau)=\tau^\kappa\Phi_1(\tau)+\tau^l\Phi_2(\tau)\log \tau+\Phi_3(\tau)$,
where $\Phi$ and $\Psi$ are smooth functions and $\Phi(0)\ne 0$.
Condition~(ii) permits  us to rewrite this in a  form
$F(\tau)=\tau\bigl(\Phi(\tau)\log \tau+\Psi(\tau)\bigr)$,
where $\Phi$ and $\Psi$ are smooth functions and $\Phi(0)\ne 0$.
If $\kappa$ is not integer, then
$\int{B(\tau)}/{\phi_0(\tau)}\,d\tau=
\tau^{l-\kappa}\psi(\tau)+C$
for some smooth function $\psi$ and some integer $l$. We have
$F(\tau)=\tau^l\psi(\tau)\phi_1(\tau)+C\phi_1(\tau)\tau^\kappa$.
Condition~(ii) implies $0<\kappa<1$.

Now let {\bf (c)} $m\le-2$, then we obtain
\begin{equation}
               \label{F-expr}
F(0)=C+\int_\sigma^0\bigl(
\tau^m\alpha_0(\tau)F(\tau)+\tau^n\beta_0(\tau)
\bigl)\,d\tau.
\end{equation}
Let us extract leading terms of both summands in the integrand;
they are given by $\mu_1(\tau)=\alpha_0(0)\tau^mF(\tau)$
and $\mu_2(\tau)=\beta_0(\tau)\tau^k$. Taking into account
condition~(ii), we have
$$
\lim_{\tau\to0}\frac{\mu_1(\tau)}{\mu_2(\tau)}=
\frac{\alpha_0(0)}{\beta_0(0)}
\lim_{\tau\to0}\frac{F(\tau)}{\tau^{k-m}}=
\cases{
\infty, & $k>m$,\cr
0, & $k\le m$.
}
$$
Thus, if $k>m$, then the leading term of the integrand in
(\ref{F-expr}) is $\mu_1(\tau)$. Since $1/\tau=o(\mu_1)$ in this case, it
follows that the integral in (\ref{F-expr}) diverges, which means that we have
no desired solutions. If $k\le n$, then the leading term is $\mu_2(\tau)$,
and the integral in (\ref{F-expr}) obviously diverges as well.
Therefore our linear differential equation is considered.

\medskip

{\bf Cubic equation}~(\ref{ce-s}) by standard substitution
$F=z+g(\tau)$, $g=-\beta/(3\tau^n)$ is reduced to the canonical
form $z^3+pz+q=0$, where $p=-(-\gamma/\tau^n+\beta/(3\tau^n))$,
$q=\delta/\tau^n-2\beta^3/(3\tau^n)^2-\beta\gamma/(3\tau^2n)$.
We use the Cardano formula
$$
z=\left(-\frac{q}{2}+\sqrt{\frac{q^2}{4}+\frac{p^3}{27}}\right)^{1/3}
+\left(-\frac{q}{2}-\sqrt{\frac{q^2}{4}+\frac{p^3}{27}}\right)^{1/3},
$$
where the branches of a cubic root are chosen in 
such a way that the product
of two summands is equal to $-p/3$.
Further analysis is based on the comparing of the orders of
$p$ and $q$ in the neighborhood of the point $y=0$.
This analysis is quite simple, although
requires certain calculations.
We omit it here, and note only, that in
the case $\frac{q^2}{4}+\frac{p^3}{27}=o(|x|^\infty)$
we have to use the Riccati equation~(\ref{riccati})
like it was done for the quadratic equation.

The following assertion summarizes all our considerations.

\begin{lemma}
Without loss of generality, the model equations {\rm(\ref{qe})},
{\rm(\ref{lde})}, {\rm(\ref{ce})}
can have only the following solutions satisfying {\rm(ii)}:
$$\begin{array}{ll}
{\rm(F1)}~F=\tau^\kappa, &{\rm(F2)}~F=\tau\log\tau,\\
{\rm(F3)}~F=\tau^{1/3}+\sigma \tau^{2/3+n},&
{\rm(F4)}~F=\tau^{2/3}+\sigma \tau^{4/3+n},
\end{array}
$$
where $0<\kappa<1$, $\sigma=\pm1$ or\/ $0$, $n$ is an integer number.
The quadratic equation {\rm(\ref{qe})} has only solution of the type {\rm (F1)}
with $\kappa=1/2$,
the linear differential equation {\rm(\ref{lde})} has solutions of the types
{\rm (F1)} and {\rm(F2)}, the cubic equation {\rm(\ref{ce})} has
solutions {\rm (F1)} with $\kappa=1/2$, {\rm(F3)} and {\rm(F4)}.
\end{lemma}

\section{Original system of equations and the singularities of selected types}

Turn back to system (\ref{e4}), and substitute obtained singular 
solutions into the system.
\begin{lemma} \label{sing}
Suppose $\kappa\not=1/2$, then up to $o(|x|^\infty)$
the following relations hold:
\begin{eqnarray}
{\rm(F1)}&:& \kappa A+ BS=0,\quad \kappa C+DS=0, \quad E=0; \label{sf1}\\
{\rm(F2)}&:& A+BS+CS=0, \quad C+DS=0, \quad A+E=0; \label{sf2}\\
{\rm(F3)}&:&\left\{
\begin{array}{l}
A+3BS+(2+3n)CS^{2n+1}+3DS^{2n+2}=0,\\
C+3DS+(2+3n)\sigma AS^n+3\sigma BS^{n+1}=0,\\
E+(1+n)\sigma C S^n+2\sigma DS^{n+1}=0;
\end{array}\right.  \label{sf3}\\
                               \label{sf4}
{\rm(F4)}&:&\left\{
\begin{array}{l}
2A+3BS+(4+3n)CS^{2n+2}+3DS^{2n+3}=0,\\
2C+DS+(4+3n)\sigma AS^n +3\sigma BS^{n+1}=0,\\
E+(2+n)\sigma CS^{n+1}+2\sigma DS^{n+2}=0.
\end{array}\right.
\end{eqnarray}
Each of these systems is consistent only if all the derivatives
of $U$ and $R$ vanish on the trajectory $X(t)$.
In other words, singular solutions of the types {\rm(F1)--(F4)}
with $\kappa\not=1/2$ do not exist.
\end{lemma}

The proof of equalities (\ref{sf1})--(\ref{sf4}) is obtained in a direct way.
The proof of inconsistency  is more delicate.

Decompose all the functions into Taylor series
and substitute these series into systems (\ref{sf1})--(\ref{sf4}),
and equate coefficients at the least powers of $x$.
The first three equations have the same form
$$
A^{(1)}=
\left(\begin{array}{c}
\langle u^{(0)},P^{(1)}\rangle R^{(0)}
+ \rho^{(0)}\langle U^{(0)},P^{(1)}\rangle \\
R^{(0)}P^{(1)}+\langle u^{(0)},P^{(1)}\rangle U^{(0)}
\end{array}\right)=0.
$$

\begin{lemma} $U^{(0)}=0$, $R^{(0)}=0$.
\end{lemma}
For the proof, let us multiply the first equation by
$\langle u^{(0)},P^{(1)}\rangle $
and subtract the second equation multiplied by $P^{(1)}$ in the sense of
the inner product and then by $\rho^{(0)}$. We obtain
$$
( \langle u^{(0)},P^{(1)}\rangle ^2-\rho^{(0)}(P^{(1)})^2) R^{(0)}=0.
$$
Since  $(P^{(1)})^2$ is a quadratic form, $\rho^{(0)}\ne 0$, and
$\langle u^{(0)},P^{(1)}\rangle $
is a linear form, it follows that the first factor
in the resulting equation does not vanish. Consequently,
$R^{(0)}=0$.
Then it readily follows from the first equation that
$\langle U^{(0)},P^{(1)}\rangle =0$, and so  $U^{(0)}=0$.
Proof is complete.

Now suppose that for some $k\ge1$ we have already
$U_j=0$, $R_j=0$, $j<k$.
Let us write out the least-order terms in the first
vector equations in systems~(\ref{sf1})--(\ref{sf4}).
They all have the same form:
\begin{eqnarray}
R^{(k)}\langle u^{(0)},P^{(1)}\rangle+
\rho^{(0)}\langle U^{(k)},P^{(1)}\rangle
+\nu S^{(2)}(\rho^{(0)}\langle\nabla, U^{(k)}\rangle+
\langle u^{(0)},\nabla\rangle R^{(k)})&=&0,\label{lo-1}\\
R^{(k)}P^{(1)}+\langle u^{(0)},P^{(1)}\rangle U^{(k)}
+\nu S^{(2)}(\nabla R^{(k)}+
\langle u^{(0)},\nabla\rangle U^{(k)})&=&0,\label{lo-2}
\end{eqnarray}
where $\nu$ is a number depending on the system.

\begin{lemma}  \label{lem8}
Under condition $u^{(0)}\not=0$ we have $U^{(k)}=0$, $R^{(k)}=0$.
\end{lemma}
This assertion for system (\ref{lo-1}) - (\ref{lo-2})
was proved in \cite{zh,full}. We omit it.

\begin{lemma}
Suppose that $R^{(j)}=0$ and $U^{(j)}=0$ for $j<k$, where $k\ge 1$. Then
$
R^{(k)}=0$.
\end{lemma}
Since $u^{(0)}=0$, equation (\ref{lo-2}) becomes
$R^{(k}P^{(1)}+\nu S^{(2)}\nabla R^{(k)}=0$.
Let us multiply it by $x$.
By the Euler identity $\langle x,\nabla R^{(k)}\rangle=k R^{(k)}$, and
$\langle P^{(1)},x\rangle=2 S^{(2)}$.
Hence $R^{(k)}(\nu k+2)=0$, which implies $R^{(k)}=0$.
The proof is complete.

Thus, to finish the proof of lemma~\ref{sing} we have to show that $U^{(k)}=0$
under the following assumption:
$u^{(0)}=0$ and $R^{(j+1)}=0$, $U^{(j)}=0$ for $j<k$, where $k\ge 1$

Let us write out the first vector equations for the least powers
of $x$ in (\ref{sf1})--(\ref{sf4}). They have the same form:
\begin{eqnarray}
                    \label{aaa}
 \langle U^{(k)},P^{(1)}\rangle+\nu S^{(2)}\langle\nabla,U^{(k)}\rangle&=&0,\\
                    \label{bbb}
R^{(k+1)}P^{(1)}+\Lambda^{(2)}U^{(k)}+\nu S^{(2)}
(
U_t^{(k)}+{}
\langle u^{(1)},\nabla\rangle U^{(k)}+{}\\
\langle U^{(k)},\nabla\rangle
u^{(1)}+
\nabla R^{(k+1)}-\omega^{(0)}T U^{(k)}
)&=&0.\nonumber
\end{eqnarray}

Now we distinguish between equations~(\ref{sf1}),~(\ref{sf2})
and equations~(\ref{sf3}),~(\ref{sf4}). We
start from equations~(\ref{sf1}) and~(\ref{sf2}).
In the following, we need the second vector equations in
systems~(\ref{sf1})--(\ref{sf2}).
We equate the coefficient of the least power
in the expansion  in powers of $x$ with zero.
This results in the equations
\begin{eqnarray}
R^{(k+1)}\langle U^{(k)},P^{(1)}\rangle+\nu S^{(2)}
\langle \nabla, R^{(k+1)} U^{(k)}\rangle&=&0
\\
\label{ddd}
\langle U^{(k)},P^{(1)}\rangle U^{(k)}
+\nu S^{(2)}\langle U^{(k)},\nabla\rangle U^{(k)}&=&0.
\end{eqnarray}

\begin{lemma} \label{11}
Equations {\rm(\ref{aaa})}, {\rm(\ref{bbb})} and {\rm(\ref{ddd})} are
compatible if and only if
$U^{(k)}=0$.
\end{lemma}

Let us multiply the (\ref{bbb})
by $P^{(1)}$ and use the expression for
$\langle U^{(k)},P^{(1)}\rangle$.
We obtain
that ${R^{(k+1)}}$ is divided by $S^{(2)}$:
$
R^{(k+1)}=\widetilde R^{(k-1)}S^{(2)},
$
Now we note that if $U^{(k)}=S^{(2)}\widetilde U^{(k-2)}$,
then the substitution $U^{(k)}$, $R^{(k+1)}$  into (\ref{aaa}), (\ref{bbb}),
(\ref{ddd})
gives the same system for $\widetilde U^{(k-2)}$, $\widetilde R^{(k-1)}$ with
$\widetilde\nu=\nu/(1+\nu)$ instead of $\nu$.
Hence without loss of generality we can assume that
$U^{(k)}$ is not divisible by $S^{(2)}$
(otherwise,  we arrive at
the original system with $k<2$ by finitely many steps).

From
equations~(\ref{aaa}), the relation $2 S^{(2)}=(P^{(1)},x)$,
and Lemma \ref{imp} we obtain
\begin{equation}
\label{uuu}
U^{(k)}=\alpha^{(k-1)}(t,x) x +\sigma^{(k-1)}(t,x) P^{(1)}_\perp,
\end{equation}
where  $\alpha^{(k-1)}=-\frac 1 2 \nu \langle \nabla,U^{(k)}\rangle $ and
$\sigma^{(k-1)}$ is a $(k-1)$-form with coefficients smooth functions of
$t$.

From~(\ref{aaa}), we
express  $\langle U^{(k)},P^{(1)}\rangle $ through
$\langle \nabla,U^{(k)}\rangle  S^{(2)}$ and
substitute into~(\ref{ddd}). Then we obtain
$$
\langle U^{(k)},\nabla\rangle U^{(k)}
=\langle \nabla,U^{(k)}\rangle U^{(k)}.
$$
Simple computations show that this equation can be rewritten as
$$
\mathop{\rm det} (\frac {\partial U^{(k)}}{\partial x})=0.
$$
It follows that the vectors  $\frac {\partial U^{(k)}_1}{\partial x}$
and   $\frac{\partial U^{(k)}_2}{\partial x}$
are collinear, i.e.,  $\frac {\partial U^{(k)}_1}{\partial
x}=\gamma_0\frac {\partial U^{(k)}_2}{\partial x}$, where  $\gamma_0(t)$
is a
smooth function. By integrating these relations with respect to $x$ and
with regard for the fact that the  $U^{(k)}$ are homogeneous
polynomials in  $x$, we readily obtain
$
U^{(k)}_2=U^{(k)}_1{\gamma_0}.
$

Let us substitute this expression into (\ref{uuu}) and multiply both sides of
the resulting relation by  ${\bf T} x$ in the sense of the inner
product; this gives
$$
U^{(k)}_1(x_2-\gamma_0 x_1)=2\sigma^{(k-1)}S^{(2)},
$$
which contradicts the assumption that $U^{(k)}$ is not divisible by
$S^{(2)}$, since  $(x_2-x_1\gamma_0)\ne 0$. This completes the proof of
Lemma \ref{11} as well as the part of Lemma \ref{sing}
pertaining to equations~(\ref{sf1}) and~(\ref{sf2}).

The consideration of the case of systems (\ref{sf3}),~(\ref{sf4})
is based on the similar ideas, but
needs more sophisticated calculations; we omit them.

\bigskip

In conclusion let us note, that the corresponding system for $F=\sqrt{\tau}$ is
$$
A+2 S B=0, \quad C+2 S D+ 2E=0,
$$
and, in contrast to the previous cases, 
it can be consistent
if $U$ and $R$ have non-vanishing derivatives.
This case is studied in~\cite{dobr2,dobr3,dobr4,zh}.

\end{document}